\documentclass[preprint,12pt]{elsarticle}
\usepackage[utf8]{inputenc}

\usepackage{graphicx}

\usepackage{amssymb}
\usepackage{amsmath}

\usepackage{hyperref}
\usepackage{algorithm}
\usepackage{algpseudocode}

\journal{Computer Physics Communications}

\begin{document}

\begin{frontmatter}



\title{PICPANTHER: A simple, concise implementation of the relativistic moment implicit Particle-in-Cell method}


\author[me]{Andreas Kempf\corref{cor1}}
\ead{ank@tp4.rub.de}
\cortext[cor1]{Corresponding author, Tel +49 234 32-23729}

\author[pk]{Patrick Kilian}
\author[ug]{Urs Ganse}
\author[fs,cs]{Cedric Schreiner}
\author[fs]{Felix Spanier}

\address[me]{Institut für theoretische Physik IV, Ruhr-Universität Bochum, Universitätsstraße 150, D-44780 Bochum}
\address[pk]{Max-Planck-Institut für Sonnensystemforschung, Justus-von-Liebig-Weg 3, D-37077 Göttingen}
\address[ug]{Department of Physics, University of Helsinki, PO Box 64, 00014 Helsinki, Finland}
\address[fs]{Centre for Space Research, North-West University, Private Bag X6001, 2520 Potchefstroom, South Africa}
\address[cs]{Institut für theoretische Physik und Astrophysik, Julius-Maximilians-Universität Würzburg, Emil-Fischer-Straße 31, D-97074 Würzburg}

\begin{abstract}
	A three-dimensional, parallelized implementation of the electromagnetic relativistic moment implicit particle-in-cell method in Cartesian geometry \cite{Noguchi_2007} is presented.
	Particular care was taken to keep the C++11 codebase simple, concise, and approachable.
	GMRES is used as a field solver and during the Newton-Krylov iteration of the particle pusher.
	Drifting Maxwellian problem setups are available while more complex simulations can be implemented easily.
	Several test runs are described and the code's numerical and computational performance is examined.
	Weak scaling on the SuperMUC system is discussed and found suitable for large-scale production runs.

	\section*{Program summary}
	\noindent
	\emph{Title of program:} PICPANTHER\\ 
	\emph{Author:} Andreas Kempf\\
	\emph{Programming language used:} C++11\\
	\emph{Computer:} Program should work on any system with a modern C++11 Compiler (e.g. g++ in GCC 4.7 and later) and MPI, HDF5 implementations\\
	\emph{Operating systems:} Linux / Unix\\
	\emph{RAM:} Variable, depending on simulation size, $\approx 2\,\mathrm{KiB}$ per cell, $56\,\mathrm{B}$ per particle\\
	\emph{Parallelization:} Parallelized using the Message Passing Interface, successfully tested on SuperMUC with good scaling behavior.\\
	\emph{Build system:} Makefiles\\
	\emph{External libraries:}\\
		Eigen3 (header files, \url{http://eigen.tuxfamily.org}, tested with versions 3.2.1, 3.2.2),\\
		MPI2 (e.g. OpenMPI, \url{http://open-mpi.org}, tested with version 1.8.1),\\
		HDF5 1.8 (\url{http://hdfgroup.org/HDF5}, tested with version 1.8.13)\\
	\emph{Nature of problem:} Kinetic simulations of collisionless plasma mostly need to resolve the smallest scales in a plasma, limiting the problem domains that can be tackled.
		The Courant–Friedrichs–Lewy condition poses further problems.
		Explicit algorithms require large amounts of computational power to cope with these restrictions.
		Implementations of implicit algorithms, on the other hand, are very complex.
		Very few implicit codes are openly available and approachable.
		Fully relativistic, three-dimensional electromagnetic implicit PiC codes in particular are rare in general.\\
	\emph{Solution method:} PICPANTHER implements the relativistic moment implicit particle-in-cell method.
		The implicit electric field equation is solved using the GMRES algorithm with operators represented as sparse matrices.
		For each particle, the implicit equation of motion is solved via a robust Newton-Krylov scheme.
		Parallelization is achieved using simple domain decomposition, resulting in good scalability.\\
	\emph{Restrictions:} PICPANTHER only allows for Euclidean geometries.
		Currently, only periodic boundary conditions are provided.\\
	\emph{Unusual features:} PICPANTHER makes use of advanced numerical techniques (GMRES, Newton-Krylov) to implicitly solve relativistic versions of the movement and field equations of a PiC code.
		It was designed to be simple and concise, using advanced C++11 language features.
		Moreover, it is parallelized and exhibits good scaling behavior.\\
	\emph{Running time:} Minutes to days, depending on problem size and CPU count\\
	\emph{Lines of code:} 1988\\
	\emph{License provisions:} CPC non-profit use license agreement\\
	\emph{CPC Library Classification:} 19.3: Collisionless Plasmas

\end{abstract}

\begin{keyword}

particle in cell \sep implicit moment \sep relativistic implicit moment \sep particle mesh \sep Newton-Krylov \sep GMRES

\end{keyword}

\end{frontmatter}



\section{Introduction} \label{sec:intro}
	Plasma physics and especially plasma astrophysics has undergone several changes in the past decades.
	For a long time fluid phenomena have dominated the research, but there was pressure to change the view of plasmas from two sides:
	Observations from high energy astrophysics demanded for an explanation of the acceleration process for very high energy charged particles \cite{Aharonian_2004}.
	On the other hand, the significant fraction of non-thermal particles in the interstellar and intergalactic medium requires a detailed understanding of instabilities caused by these particles.

	Kinetic plasma descriptions, required to understand and model the non-thermal components of a plasma, come in a wide variety.
	A full description of kinetic plasmas requires the solution of the Boltzmann equation coupled with Maxwell's equations and a complicated collision operator.
	Finding an analytical or even a numerical solution for this problem set seems impossible at the moment.
	One usually resorts to fully non-collisional plasmas, which can be described by the Vlasov-Maxwell system.
	Still, this simplification yields a six-dimensional phase space problem that is too memory intensive to be solved for general three-dimensional problems.

	The Particle-in-Cell (PiC) method \cite{birdsall_langdon, hockney_eastwood} has established itself as a state of the art method for solving problems in kinetic plasma physics.
	It is a compromise between direct particle interaction, i.e. molecular dynamics or N-body codes, and field only methods (Vlasov codes \cite{buechner}).
	The main advantages of the PiC method are that their memory consumption increases linearly with the simulated volume and that the runtime is only of order N.
	They are also very suitable for the use of large multi-processor systems.
	Their main disadvantages, on the other hand, are high noise levels and high computational requirements due to the operation on the shortest time and length scales.
	Implicit methods like the one used in our code can alleviate the computational burden by allowing for larger timesteps and cell sizes, making larger scale simulations a possibility.

	As has been discussed in the literature \cite{birdsall_langdon, hockney_eastwood}, the general outline of PiC codes is very simple and can be summarized in four stages:
	Current assignment, field propagation, force calculation and particle movement.
	A similar simplicity was aimed for during the development of this code.
	Particular care is taken to keep complexity low by adhering to modern C++ language features and paradigms.
	Despite the numerical techniques being highly advanced, the code itself is, therefore, easy to understand.
	The result is PICPANTHER (Parallel Implicit Concise PiC Allowing Non-THermal Electromagnetic Relativity) a scalable, implicit, three-dimensional PiC code implementing the relativistic moment method \cite{Noguchi_2007} with GMRES \cite{gmres} and Newton-Krylov \cite{Kelley_2003} solvers for fields and particles respectively.

\section{Definitions} \label{sec:definitions}
	An extensive derivation of the relativistic moment implicit particle-in-cell method in Cartesian geometry is given in \cite{Noguchi_2007}.
	The quantities from \cite{Noguchi_2007} that are relevant to this paper are summarized in the following paragraphs with their derivations omitted here for brevity.
	In this section, particle positions are denoted by $\vec{x}$ and particle velocities by $\vec{v}$ or $\vec{u} = \gamma \vec{v}$ with the gamma factor $\gamma$.
	SI units are used throughout this paper and the code.

	Firstly, a "magnetic field rotation tensor", often encountered in implicit particle-in-cell schemes \cite{avanti, petdav}, needs to be calculated for each individual particle:
	\begin{equation} \label{eq:mag_tensor}
		\hat{\alpha} = \frac{1}{D}
		\left(
			\hat{I}
			- \beta' \hat{T}[{\vec{B}^{\,n}}]
			+ \beta'^{\,2} \vec{B}^{\,n} \otimes \vec{B}^{\,n}
		\right)
	\end{equation}
	where
	\begin{align}
		\beta &= \frac{q \Delta t}{2 m}
		\mbox,\\
		\Gamma &= \frac{\beta}{c^2} \vec{E}^{\,n} \cdot \vec{v} + \gamma
		\mbox,\\
		\beta' &= \frac{\beta}{\Gamma}
		\mbox,\\
		D &= \Gamma (1 + \beta'^{\,2} \vec{B} \cdot \vec{B})
		\mbox,
	\end{align}
	with the particle charge $q$ and mass $m$, the simulation timestep $\Delta t$, and $\hat{T}[\vec{B}^{\,n}]$ being the skew-symmetric matrix representing the cross-product of $\vec{B}^{\,n}$ with an arbitrary vector.
	This tensor is a rotation transformation due to the magnetic field.
	In a non-relativistic implementation, $\hat{\alpha}$ does not need to be calculated for each particle individually.
	It can be computed directly on the grid instead, with it being defined per species due to the gamma factor being unity.

	The source terms required for the solution of Maxwell's equations are the charge density $\rho$, the current density $\vec{j}$, the pressure tensor $\hat{\Pi}$, and the dielectric tensor $\hat{\mu}$:
	\begin{align}
		\label{eq:sources-r}
		\rho (\vec{x}_\mathrm{g})
		&=W(\vec{x}-\vec{x}_\mathrm{g}) \cdot
		\sum
		\frac{1}{\Delta x^3}&
		q
		\mbox,\\
		\label{eq:sources-j}
		\vec{j} (\vec{x}_\mathrm{g})
		&=W(\vec{x}-\vec{x}_\mathrm{g}) \cdot
		\sum
		\frac{1}{\Delta x^3}&
		q ( \hat{\alpha} \cdot \vec{u} )
		\mbox,\\
		\label{eq:sources-p}
		\hat{\Pi} (\vec{x}_\mathrm{g})
		&=W(\vec{x}-\vec{x}_\mathrm{g}) \cdot
		\sum
		\frac{1}{\Delta x^3}&
		q ( \hat{\alpha} \cdot \vec{u} ) \otimes ( \hat{\alpha} \cdot \vec{u} )
		\mbox,\\
		\label{eq:sources-m}
		- \hat{\mu} (\vec{x}_\mathrm{g})
		&=W(\vec{x}-\vec{x}_\mathrm{g}) \cdot
		\sum
		\frac{1}{\Delta x^3}&
		\frac{\Theta (q \Delta t)^2}{2 \varepsilon_0 m}
		\hat{\alpha}
		\mbox.
	\end{align}
	The dimensionless filtering parameter $\Theta$ is introduced below.
	The edge length of a single cubic cell is represented by $\Delta x$ and the vacuum permittivity by $\varepsilon_0$.
	These quantities are deposited in the standard particle-in-cell manner, summing over all macro-particles weighted with an interpolation function $W$.
	Usually, $W(\vec{x})$ is a product of splines $s_\mathrm{m}(x)$, $s_\mathrm{m}(y)$, $s_\mathrm{m}(z)$ of some order $m$.
	Choosing a suitable $m$ is necessarily a compromise between high computational performance (low order) and low numerical noise (high order).

	The terms $\rho$, $\vec{j}$, $\vec{\Pi}$ make up the actual charge density $\widetilde{\rho}$ to be used in the calculation of the electric field,
	\begin{equation}
		\widetilde{\rho} = \rho - (\Theta \Delta t) \nabla \cdot
		\left(
			\vec{j} - \frac{\Delta t}{2} \nabla \cdot \hat{\Pi}
		\right)
		\mbox.
	\end{equation}
	After several algebraic manipulations, an implicit expression for the electric field $\vec{E}^{\,n+\Theta}$ is obtained,
	\begin{multline} \label{eq:e_field}
		(c \Theta \Delta t)^2
		\left(
			- \nabla^2 \vec{E}^{\,n+\Theta}
			+ \nabla \nabla \cdot
			\left(
				\hat{\mu} \cdot \vec{E}^{\,n+\Theta}
			\right)
		\right)
		+ \left(\hat{I} - \hat{\mu} \right) \cdot \vec{E}^{\,n+\Theta}
		=\\ \vec{E}^{\,n} -
		c^2 \Theta \Delta t
		\left(
			\mu_0 \vec{j}
			-\frac{\mu_0 \Delta t}{2} \nabla \cdot \hat{\Pi}
			-\nabla \times \vec{B}^{\,n}
		\right)
		- (c \Theta \Delta t)^2 \nabla \frac{\widetilde{\rho}}{\varepsilon_0}
		\mbox,
	\end{multline}
	with $\Theta$ being a filtering parameter, $\Theta \in [0.5, 1.0]$, that controls the propagation of electromagnetic waves.
	Choosing $\Theta = 1.0$ almost completely removes the electromagnetic mode from the simulation.
	By contrast, $\Theta = 0.5$ results in a second order correct timestepping, the EM mode staying intact.
	$\vec{E}^{\,n+1}$ is calculated by linear extrapolation.
	\begin{equation} \label{eq:interpolation}
		\vec{E}^{\,n+1} = \frac{\vec{E}^{\,n+\Theta} + (\Theta - 1) \vec{E}^{\,n}}{\Theta}
		\mbox.
	\end{equation}

	The magnetic field is updated according to Faraday's law.
	\begin{equation} \label{eq:b_field}
		\vec{B}^{\,n+1} = \vec{B}^{\,n}
		- \Delta t
		\left(
			\nabla \times \vec{E}^{\,n+\Theta}
		\right)
		\mbox.
	\end{equation}

	From the equations of motion, a nonlinear residual equation is derived for the update of a single particle,
	\begin{equation} \label{eq:part}
		\vec{r} (\vec{u}^{\,n+1})
		= \frac{\vec{u}^{\,n+1} - \vec{u}^{\,n}}{\Delta t}
		- \frac{q}{m}
		\left(
			\vec{E}^{\,n+\Theta}(\vec{x}^{\,n+1/2})
			+ \frac{\vec{u}^{\,n+1} + \vec{u}^{\,n}}{\gamma^{n+1} + \gamma^n}
			\times \vec{B}^{\,n} (\vec{x}^{\,n+1/2})
		\right)
		\mbox.
	\end{equation}
	Of note here is the fact that the electromagnetic fields in this equation are required at the mid-orbit position of the particle $\vec{x}^{\,n+1/2}$.
	Consequently, the equation needs to be solved iteratively.
	During and after the solution of equation \eqref{eq:part}, the particle's location is updated according to
	\begin{equation} \label{eq:part-loc}
		\vec{x}^{\,n+1} = \vec{x}^{\,n} + \Delta t \frac{\vec{u}^{\,n+1} + \vec{u}^{\,n}}{\gamma^{n+1} + \gamma^n}
		\mbox.
	\end{equation}

	The definitions above are not exactly equal to the ones in \cite{Noguchi_2007} due to small oversights being corrected.
	The expression for $\hat{\mu}$ contains an additional factor $q$.
	In equation \eqref{eq:e_field}, a factor of $\mu_0$ was added.
	Finally, signs where changed in equations \eqref{eq:e_field} and \eqref{eq:b_field}.

	Being a more general algorithm, the scheme outlined above can be modified easily to yield a simpler, non-relativistic version \cite{ipic3d}.

\section{Discretization} \label{sec:discretization}
	First order finite differences are employed to obtain the spatial derivatives needed for the vector operators.
	\begin{equation}
		\left( \frac{\partial A}{\partial x} \right) _\mathrm{i+1/2,j,k} = \frac{A_\mathrm{i+1,j,k} - A_\mathrm{i,j,k}}{\Delta x}
		\mbox.
	\end{equation}
	Where required, missing values are taken as an average:
	\begin{equation}
		A_\mathrm{i,j+1/2,k+1/2} = \frac{A_\mathrm{i,j+1,k} + A_\mathrm{i,j,k} + A_\mathrm{i,j,k+1} + A_\mathrm{i,j+1,k+1}}{4}
		\mbox.
	\end{equation}
	Two possible distributions of grid quantities were tested.
	First, the lattice arrangement is chosen such that the electric field $\vec{E}$ along with $\vec{j}$ and $\hat{\mu}$ are located at cell nodes.
	The magnetic field $\vec{B}$, $\rho$, and $\hat{\Pi}$ are defined at cell centers.
	Second, the quantities are arranged such that the electric and magnetic fields constitute a standard Yee-lattice \cite{yee}.
	This can be realized by depositing the components of $\vec{E}$ and $\vec{j}$ on cell edges parallel to the component direction and the components of $\vec{B}$ on cell faces perpendicular to the component direction.
	The charge density $\rho$ is located at cell nodes, the components of $\hat{\Pi}$ such that their divergence is evaluated at cell edges and $\hat{\mu}$ such that $\hat{\mu} \cdot \vec{E}$ results in a vector stored like $\vec{E}$.
	Expressing the gradient, divergence, and curl at cell nodes and centers is straightforward in both cases, if tedious.
	Since the discretized operators are matrices, two operators can be combined easily by a simple matrix multiplication.

	For the code, the Yee arrangement was chosen as default.
	The resulting matrices are simpler and contain fewer elements.
	Moreover, the dispersion relations obtained below are clearer with modes being more pronounced.
	The cell-centered / node-centered scheme is still available.
	Unfortunately, switching schemes requires the replacement of three source code files.

\section{Code} \label{sec:code}
	As described in section \ref{sec:discretization}, the discrete operators are simple matrices.
	Like with most discretizations of partial differential equations, the resulting matrices are sparse.
	By using a modern sparse matrix library, the equations can be written naturally in a very compact and natural manner.
	Here, the Eigen library \cite{eigen} was utilized.
	However, alternative libraries like Armadillo \cite{armadillo} could be substituted effortlessly, provided a suitable sparse matrix solver is available.
	As an example, the C++ code calculating equation \eqref{eq:b_field} can be written
	\begin{verbatim}
	B -= dt * curl_E * E;
	\end{verbatim}
	using a sparse matrix \verb+curl_E+ (Yee-scheme, \verb+curl_center+ in other scheme) that was prepared at the beginning of the simulation run according to the discretization rules outlined above.
	Combinations of operators are similarly just multiplications in the code.

	Currently, only periodic boundary conditions are implemented.
	Taking the simulation volume to be a three-dimensional box of size $N_\mathrm{x} \times N_\mathrm{y} \times N_\mathrm{z}$, the scalar, vector, and tensor fields become regular vectors of size $N_\mathrm{x} N_\mathrm{y} N_\mathrm{z}$, $3 N_\mathrm{x} N_\mathrm{y} N_\mathrm{z}$, $9 N_\mathrm{x} N_\mathrm{y} N_\mathrm{z}$, respectively.
	Accordingly, equation \eqref{eq:e_field} is a $3 N_\mathrm{x} N_\mathrm{y} N_\mathrm{z} \times 3 N_\mathrm{x} N_\mathrm{y} N_\mathrm{z}$ square sparse matrix operating on $\vec{E}^{\,n+1}$ with the right-hand side also being a vector of size $3 N_\mathrm{x} N_\mathrm{y} N_\mathrm{z}$.
	Since this is a straightforward matrix equation, any solver capable of solving a sparse, non-symmetric, square matrix system can be employed to solve for the advanced electric field.
	As in \cite{Noguchi_2007}, the generalized minimal residual method (GMRES) \cite{gmres} was chosen for this task.
	GMRES is implemented as an unsupported module in Eigen3.
	Switching to a different solver (e.g. the biconjugate gradient stabilized method BiCGSTAB \cite{bicgstab}) is possible, although no such alternatives have been explored yet.

	Solving the nonlinear equation \eqref{eq:part} for the time-advanced velocity is achieved using a Newton-Krylov method.
	A regular GMRES algorithm \cite{Kelley_2003, gmres_imp} is modified by replacing the matrix-vector-products with scaled numerical differences as described in \cite{Knoll} to yield the inner loop of the Newton iteration.
	Moreover, a basic line search method is employed to improve convergence \cite{Kelley_2003}.

	Additionally, the code can operate in parallel with the available processors each assigned a box of about equal size.
	Equation \eqref{eq:e_field} is iterated via a regular Schwarz domain decomposition method \cite{schwarz} using ghost cells.
	Data transfer between processors is mediated by the message passing interface (MPI).
	Parallel output, provided by the HDF5 library, is available for particles and grid quantities.

	Performance is an important feature of a particle in cell code.
	Currently, with only little optimization work done, the code performs about an order of magnitude fewer particle updates per second when compared to an optimized code like ACRONYM \cite{Kilian_2011}.
	Due to many force interpolations being performed during the Newton-Krylov iteration using a TSC form factor, performance necessarily suffers when compared to an explicit scheme.
	In fact, even with few particles per cell the computational time is mostly spent in the particle solver.
	This does, of course, also depend on the number of Newton iterations and the tolerance specified for the residual reduction.
	Taking this fact into account, the code performance seems adequate, especially considering the benefits of an implicit scheme.
	Further optimization is certainly always desirable.

	Memory requirements are highly dependent on the particle and cell count.
	Each cell requires about 40 64 bit floating point values being stored on the grid.
	Additionally, more than 100 values per cell are needed for the sparse matrix operators (Yee, more for the other scheme).
	This amounts to approximately $2 \, \mathrm{KiB}$ of storage per cell (including ghost cells).
	A single particle consists of seven 64 bit values resulting in $56 \, \mathrm{B}$ per particle.

	The code is written in pure C++ while adhering to idioms like RAII (Resource Acquisition Is Initialization), avoiding pointers and explicit memory management.
	These principles allow for very compact code when coupled with modern C++11 language and standard library features.
	Taking advantage of a matrix library like Eigen also significantly benefits compactness and simplicity.
	As a consequence, the full codebase is under 2000 lines of code, as counted by the CLOC \cite{cloc} utility (duplicated code for the two discretization schemes was ignored).
	MPI communication, HDF output, and the operator setup necessarily make up a large part of the code but are relatively straightforward.

\begin{figure}
		\centering
		\includegraphics[width=\textwidth]{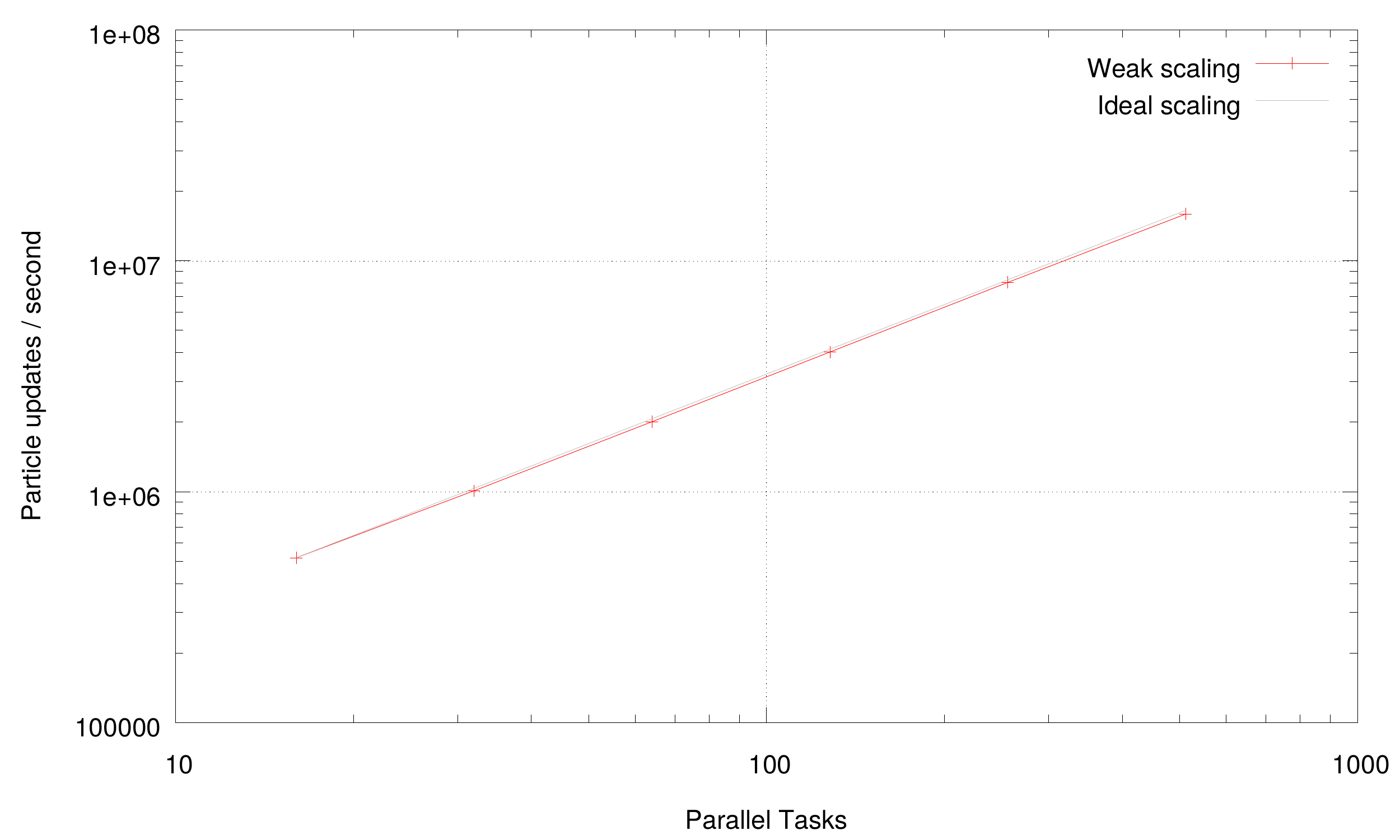}
		\caption{
			Weak scaling on 1, 2, 4, 8, 16, and 32 SuperMUC nodes.
			$48 \times 48 \times 32$ cells per core were arranged along the z-axis with 20 particles per cell.
		}
		\label{fig:benchmark}
	\end{figure}

\begin{figure}
		\centering
		\includegraphics[width=\textwidth]{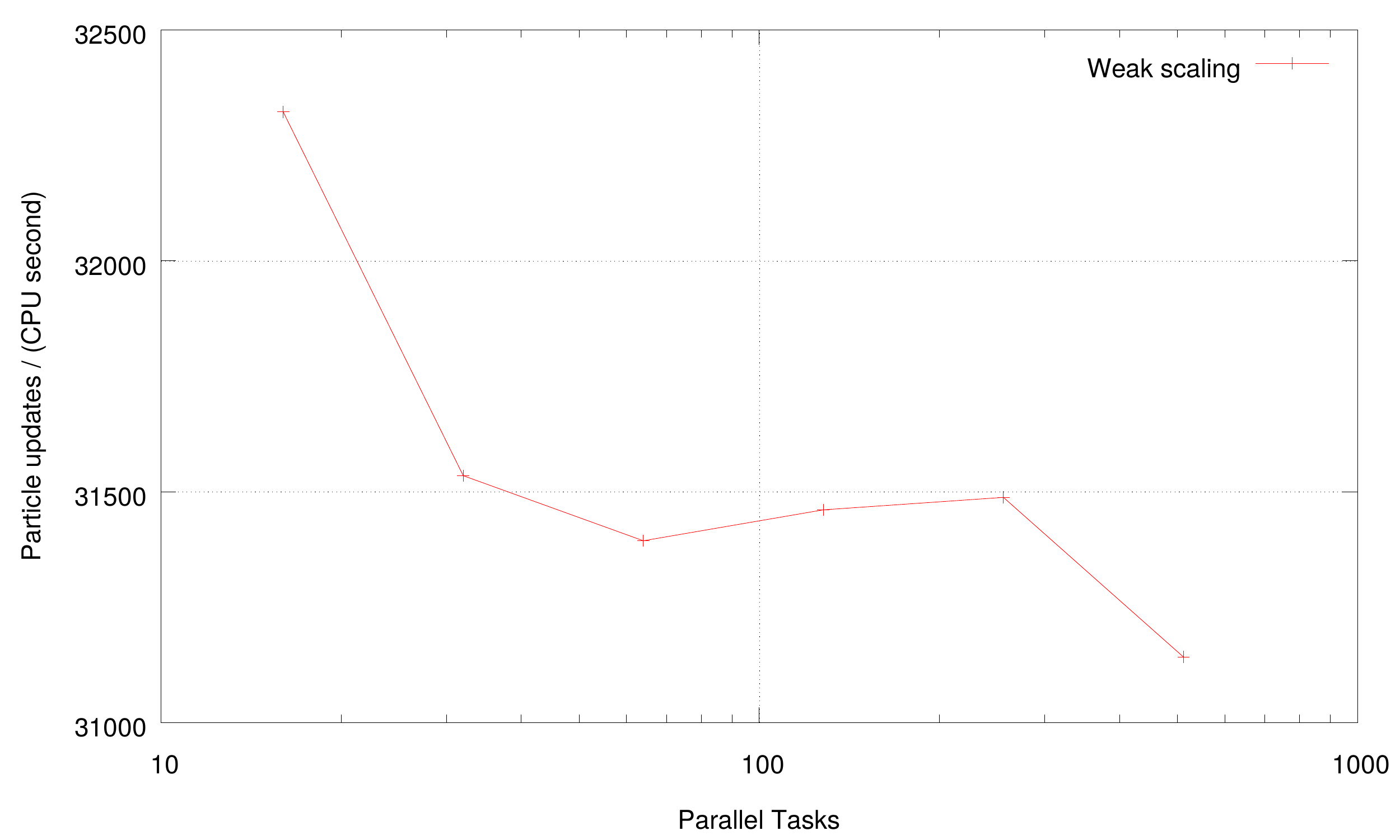}
		\caption{
			Loss of efficiency in the benchmark runs of fig. \ref{fig:benchmark}.
			The behavior between 32 and 256 cores is reproducible and seems to be a result of SuperMUC topology and task placement on the machine.
		}
		\label{fig:benchmark-lin-log}
	\end{figure}

	Successful test runs on the SuperMUC supercomputing system were performed on up to 512 cores (32 nodes).
	Weak scaling performance is plotted in fig. \ref{fig:benchmark} and \ref{fig:benchmark-lin-log}.
	The total number of particle updates is a product of the cell count, particles per cell, and number of timesteps.
	For the plots, this number is divided by the total wall clock time or total CPU-time respectively.
	Therefore, it is a performance measure of the complete update cycle.
	With a PiC code's computing requirements being mostly determined by the total number of particles in the simulation volume, this quantity allows a direct comparison between different codes.
	In this case, the number of field iterations is fixed and the number of Newton iterations per particle is approximately constant over the simulation.
	During the benchmark runs with 20 particles per cell, the source term accumulation, field calculation, and particle update substeps each take up roughly $20\%$, $15\%$, $65\%$ of the time, respectively.
	Only small deviations from ideal scaling, extrapolated from 16 cores (1 node), are visible.


\section{Implementation} \label{sec:implementation}
	In this section, a description of the steps executed during a production run is given.

\subsection{Initialization}
	At first, a configuration file called \verb+config+ in the working directory is parsed.
	Its format will be described in a different section below.
	Provided in the configuration file are physical parameters like the electron plasma frequency, simulation details like the desired number of timesteps and cells in each direction, and program settings like the number of MPI processes distributed along each direction.
	Next, the MPI communication is set up and the simulation volume is divided among processes.
	Furthermore, the HDF5 file \verb+output.h5+ and a plain text file \verb+energy_output.dat+ is created by rank 0.
	Important parameters like the length of a timestep and the size of a cell are calculated and the layout of the sparse matrix operators is prepared.
	This concludes the programmatic setup sequence.

	With the parameters given in the configuration file, the magnetic and electric field configuration and particle distributions are initialized.
	For simplicity, a simple standard setup for a drifting Maxwellian particle distribution in a constant background magnetic field is available.
	Particles are created using data from a pseudo-random number generator with a given seed, allowing for repeatable simulation runs.
	Standard C++11 facilities are used to draw normally distributed velocity components with a given variance for each particle.
	Similarly, a uniform distribution provides the particle location within a given cell.

	The physical setup ends with a first output of the data.

\subsection{Timestep}
\begin{algorithm}
	\begin{algorithmic}[1]
		\caption{Update cycle}
		\label{alg:timestep}
		\Repeat
		\ForAll{particles}
			\State {collect \verb+charge_density+}
			\Comment{eq. \eqref{eq:sources-r}}
			\State {collect \verb+current_density+}
			\Comment{eq. \eqref{eq:sources-j}}
			\State {collect \verb+pressure+}
			\Comment{eq. \eqref{eq:sources-p}}
			\State {collect \verb+dielectric+}
			\Comment{eq. \eqref{eq:sources-m}}
		\EndFor
		\State {distribute source terms among processes}
		\State {\verb+E_old+ $\gets$ \verb+E+}
		\State {\verb+B_old+ $\gets$ \verb+B+}
		\State {prepare \verb+operator+ matrix}
		\Comment{eq. \eqref{eq:e_field}}
		\State {prepare \verb+right-hand-side+ vector}
		\Comment{eq. \eqref{eq:e_field}}
		\For{$i \gets 0, \mathrm{max\_iters}$}
			\State {\verb+E+ $\gets$ GMRES(\verb+operator+, \verb+right-hand-side+, guess=\verb+E+)}
			\State {distribute \verb+E+ among processes}
		\EndFor
		\State {\verb+B+ $\gets$ \verb+B+ - \verb+dt+ * \verb+curl_E+ * \verb+E+}
		\Comment{eq. \eqref{eq:b_field}}
		\ForAll{particles}
			\While{residual $>$ tolerance}
				\State {\verb+u_new+, residual $\gets$ NK(\verb+E+, \verb+B_old+, \verb+u_new+)}
				\Comment{eq. \eqref{eq:part}}
			\EndWhile
			\State {\verb+x+ $\gets$ \verb+x+ + \verb+dt+ * (\verb+u+ + \verb+u_new+) / (\verb+g+ + \verb+g_new+)}
			\Comment{eq. \eqref{eq:part-loc}}
		\EndFor
		\State {distribute particles among processes}
		\State {\verb+E+ $\gets$ (\verb+E+ - (1-\verb+THETA+) * \verb+E_old+)/\verb+THETA+}
		\Comment{eq. \eqref{eq:interpolation}}
		\Until{simulation finished}
	\end{algorithmic}
\end{algorithm}

	In this section, a single simulation timestep is described in detail.
	Algorithm \ref{alg:timestep} roughly outlines the simulation cycle in pseudocode.

	Initially, the source grid quantities for pressure, dielectric tensor, charge density and current density are set to zero.
	For each particle in the simulation volume, the field quantities at its current position are interpolated and its contribution to the source terms calculated.
	This contribution is then deposited onto the grid according to the chosen grid scheme.
	After all the terms are summed up, the values stored on ghost cells are communicated between neighboring processes.
	At this point, the quantities $\rho$, $\vec{j}$, $\hat{\Pi}$, and $\hat{\mu}$ are known on the grid.
	Additionally, a sparse matrix operator is created for the dielectric tensor and a helper variable for $\nabla \cdot \hat{\Pi}$ is stored.
	An optional smoothing step is available for the charge density.
	This step can improve energy conservation and reduce noise in some cases.

	Given the source terms accumulated above, Maxwell's equations can be solved implicitly.
	Fields from the previous timestep are renamed as they are still needed after the update.
	Using the differential operators prepared during setup, a GMRES solver is initialized for the electric field update.
	Moreover, a vector describing the right-hand side is calculated from the source terms.
	Taking the current electric field values as an initial guess, a GMRES step is taken.
	After the GMRES step, the border values of the electric field are exchanged among neighboring processes.
	This refinement and distribution of the electric field is repeated for a fixed number of iterations since the solution seems to converge rapidly.
	For large timesteps and cell sizes, more iterations might be required here.
	Having thus evaluated $\vec{E}^{\,n+\Theta}$, the magnetic field can be updated and synchronized among the processes.
	Due to the filtering parameter $\Theta$, a linear extrapolation of $\vec{E}^{\,n+1}$ is performed later (see below).

	Thus, with $\vec{E}^{\,n+\Theta}$ and $\vec{B}^{\,n+1}$ known, the particle velocities and positions need to be updated.
	For each particle, a solution for its time-advanced velocity is calculated in a Newton-Krylov procedure.
	It uses evaluations of the residual to continuously prepare numerical approximations to the Jacobian matrix.
	The Jacobian matrix is used in a GMRES iteration to obtain a new approximation for a Newton step.
	This step is then applied using a simple line search algorithm and a new GMRES iteration is performed.
	After a set number of steps, the particle is moved with the velocity minimizing the residual.
	The residual itself is evaluated by interpolating the forces acting at the particle's location and solving equation \eqref{eq:part}.

	Unfortunately, Newton-Krylov procedures are not guaranteed to converge.
	Similarly, catastrophic failures of the solver cannot be ruled out in a production run with billions or even trillions of particle updates.
	Consequently, velocities returned by the solver are checked for validity.
	If an invalid velocity (NaN) is encountered or the Newton iteration fails to converge, the particle is moved using its previous value instead and a warning is printed.
	A few isolated corrections should not greatly influence a sufficiently large simulation.
	Extensive testing revealed that the Newton-Krylov algorithm is sensitive to floating point round-off issues.
	For this reason, the code performs better on processors implementing a fused multiply–add instruction (e.g. FMA4 on AMD Bulldozer 2011 and later, FMA3 on Intel Haswell 2013 and later), assuming it is supported by the compiler.
	While measures where taken to mitigate these issues, simulations on processors listed above show faster convergence and fewer catastrophic iteration failures (see also \cite{newton_stability}).
	Since fewer iterations are needed in that case, performance also increases.
	Taking the above into account, the particle solver should always be monitored and its parameters adjusted if necessary.

	Concluding the physical part of a timestep, the electric field is extrapolated to yield the proper values needed for the next iteration as per equation \eqref{eq:interpolation}.

	Diagnostic output of the total energy in a simulation is written to the text file \verb+energy_output.dat+ after each step by rank 0.
	Field and particle output is optional after each step.
	During field output, rank 0 first creates a skeleton group structure in \verb+output.h5+.
	Then, all processes write their field data collectively.
	For particle output, a new folder is created.
	Each processor writes its particle data to a separate file in the folder using the packet table interface.

\section{Usage}

\subsection{Compilation and Execution}
	A self-explanatory makefile is provided.
	Ideally, the HDF5 environment is set up properly, so that the \verb+h5c+++ command wraps \verb+mpiCC+ (or equivalent).
	That way, all the necessary libraries and header files for MPI and HDF are taken care of automatically.
	Eigen3 headers are assumed to reside in \verb+/usr/include/eigen3+.

	The resulting binary is executed via \verb+mpiexec+ (or equivalent).
	The number of processes needs to be specified via the standard methods of the MPI implementation (e.g. OpenMPI: \verb+mpiexec -np 4 .../imp+ for four processes).

\subsection{Configuration}
	Configuration data for the code is read from a file \verb+config+ in the working directory.
	Its format is one item \verb+key=value+ per line without any other white space.
	The key is a character string and the value is a double precision floating point value.
	Available configuration options are listed in table \ref{tab:config} and an example file is provided with the code.

	\begin{table}
		\centering
		\begin{tabular}{l|c|p{80mm}}
			Key & Unit & Description\\
			\hline
			\verb+seed+ & - & Seed for the random number generator\\
			\verb+total_steps+ & - & Number of timesteps to be executed\\
			\verb+N_x+, (\verb+y+, \verb+z+) & - & Number of cells in x, y, z direction\\
			\verb+procs_x+ (\verb+y+, \verb+z+) & - & Number of processors distributed along the x, y, z axes\\
			\verb+plasma_freq+ & $\mathrm{rad/s}$ & Electron plasma frequency\\
			\verb+mp_over_me+ & - & Proton mass divided by electron mass\\
			\verb+width_bg+ & c & Width of a velocity component in thermal distribution (background)\\
			\verb+width_jet+ & c & Width of a velocity component in thermal distribution (jet)\\
			\verb+num_*_bg+ (\verb+e+, \verb+P+, \verb+p+) & - & Number of macro- $e^-$, $p$, $e^+$ per cell (background)\\
			\verb+num_*_jet+ (\verb+e+, \verb+P+, \verb+p+) & - & Number of macro- $e^-$, $p$, $e^+$ per cell (jet)\\
			\verb+v_*_bg+ (\verb+x+, \verb+y+, \verb+z+) & c & Particle drift velocity in x, y, z direction (background)\\
			\verb+v_*_jet+ (\verb+x+, \verb+y+, \verb+z+) & c & Particle drift velocity in x, y, z direction (jet)\\
			\verb+B0_x+ (\verb+y+, \verb+z+) & T & Initial magnetic field  in x, y, z direction\\
			\verb+rescale_dx+ & - & Cell size $dx = \lambda_{\mathrm{D}}/\sqrt{2}$ will be divided by this factor\\
			\verb+rescale_dt+ & - & Time step $dt = dx / (\sqrt{3} c)$ will be divided by this factor\\
			\verb+theta+ & - & Filtering parameter $\Theta \in [0.5, 1.0]$\\
			\verb+out_p+ & - & Particle data is written every ... steps\\
			\verb+out_q+ & - & $\rho$ is written every ... steps\\
			\verb+out_j_x+ (\verb+y+, \verb+z+) & - & $j_\mathrm{x/y/z}$ is written every ... steps\\
			\verb+out_E_x+ (\verb+y+, \verb+z+) & - & $E_\mathrm{x/y/z}$ is written every ... steps\\
			\verb+out_B_x+ (\verb+y+, \verb+z+) & - & $B_\mathrm{x/y/z}$ is written every ... steps\\
			\verb+smooth_charge+ & - & A binomial filter will be applied to $\rho$ if this is $>= 1$
		\end{tabular}
		\caption{Configuration options.}
		\label{tab:config}
	\end{table}

\subsection{Creating a new setup}
	Creating an initialization procedure sufficiently powerful and general to cover most use cases using only a configuration file is an awkward procedure.
	Additional parsing procedures, rules and exceptions can quickly lead to ballooning complexity and in many cases, the code needs to be edited anyway.
	Consequently, the present code only provides a setup using up to two homogeneous drifting Maxwellian distributions.
	If a more complex setup is needed, it is simple to create a new configuration by editing the source code directly.
	Since the existing files are commented, they should be used as a reference.

	The functions \verb+init_...+ are executed at the start of a simulation in the order parameters, sources, fields, and particles.
	Cell size \verb+dx+, \verb+Particle::dx+ (static variable) timestep length \verb+dt+, \verb+Particle::dt+ (static variable) need to be set in one of these functions.
	Moreover, the particle data (charge $q$, mass $m$, $\beta/dt$, number density $n$) contained in the \verb+Particle::p+ array should be defined properly for electrons (index 0), protons (index 1), and positrons (index 2).

	By looping from 0 to Nx, 0 to Ny, 0 to Nz, every cell in the simulation volume can be indexed.
	For access to vector field components (e.q. $E_\mathrm{x}$), the helper function \verb+vindg+ is used with indices x,y,z and the component c (0=x,1=y,2=z).
	Scalar fields (if needed) are accessed with \verb+sindg+ and indices x,y,z.
	Both helper functions map the three cell indices to a single integer index, taking into account ghost cells automatically.

	Particles are simply added to the \verb+parts+ vector.
	The \verb+Tag+ union is provided to distinguish particles along with their properties.
	It is important to set the \verb+flav+ field to one of the \verb+FLAVOR_...+ values since this determines the properties of the particle.
	A field \verb+population+ is provided in addition to two integers \verb+id_in_cell+, \verb+start_cell_index+ allowing for unique particle IDs, if needed.
	Such an ID can be created by calculating the unique global index of the current cell and storing that value in \verb+start_cell_index+.
	Different particles in the same cell receive incrementing values in \verb+id_in_cell+.
	Population identifiers (\verb+POP_...+) are optional and serve to differentiate populations of the same particle type (e.g. jet or background).

	New variables needed during initialization may be put into the \verb+Simulation+ class in header \verb+simulation.h+.
	If other particle types or populations are needed as well, new \verb+FLAVOR_...+ and / or \verb+POP_...+ entries should be added in file \verb+particle.h+.
	The \verb+FLAVOR_...+ constants serve as indices into the properties array.
	Therefore, the property array needs to be large enough to contain the particle properties.
	In \verb+particle.cpp+, the \verb+Particle::p[]+ initialization needs to be extended accordingly.
	As mentioned, the final particle properties need to be set up in one of the \verb+init+ functions.

	Parameters read from the configuration file are accessible through the \verb+parameters+ hash table and indexed using the key used in the file.

\subsection{Processing output}
	In \verb+output.h5+ field data for the simulation is stored.
	For each output timestep, a new group is created, e.g. \verb+Timestep_0+.
	In this group, new groups for each field vector are created, e.g. \verb+E+.
	Finally, datasets for its components are created in this group, e.g. \verb+E0+.
	The full path for $E_x$ at timestep 0 would be \verb+/Timestep_0/E/E0+, accordingly.
	The program code includes a short Python script that creates a dispersion plot from an output file.

	Similarly, particles at timestep 0 are stored in a folder called \verb+particles_0+.
	Each CPU creates its own file named \verb+cpu_0+, for example.
	The files contain a standard HDF5 packet table consisting of the complete particle data ($\vec{r}$, $\vec{u}$, ID).
	By parsing the ID field, particle details like the type can be extracted.
	A Python example script is provided for the creation of a histogram from particle output.

	Convenient Python modules for the handling of the output datasets are NumPy \cite{numpy}, SciPy \cite{scipy}, PyTables \cite{pytables} and matplotlib \cite{matplotlib}.
	Many utilities can handle HDF5 natively so Python need not be used for post-processing.
	VisIt, for example, can plot the field output without further preparation \cite{visit}.

\section{Simulation setup}

The capabilities of the PiC code presented here shall be highlighted with a few examples.

\subsection{Wave dispersion}
	One important aspect of PiC codes is the interaction of particles and fields.
	A simple way to test whether these interactions are reflected properly is the generation and propagation of plasma waves.
	Although they are relying on the complex interplay of particles and waves, they are well understood in terms of theory.
	It is therefore possible to compare the waves' properties to known expressions \cite{stix_1992}.

	A thermal magnetized plasma generated randomly will always contain several wave modes with dispersion relations that can be evaluated analytically.
	This simple setup is realized by creating a homogeneous distribution of electrons and protons with Maxwellian velocity distributions.
	Depending on the wave mode, the runtime of the simulation can be chosen in order to resolve the lowest frequencies of interest.
	Similarly, the number of cells in each direction is chosen according to the largest wavelength to be resolved along the respective axis.

	The relevant parameters used in these simulations are summarized in table \ref{tab:parameters-waves}.
	Filtering parameters $\Theta=0.5$ and $\Theta=1.0$ were tested.
	In the former simulation the charge density was smoothed once using a binomial filter.
	No further smoothing was performed in these simulations.

	To analyze the dispersion relation along an axis, the simulation fields are integrated over the perpendicular directions and Fourier-transformed in space and time.
	Thereby, $\omega(k_\mathrm{i})$ plots can be obtained along an axis $\mathrm{i}$ for all quantities represented on the computational grid.
	For comparison, an identical simulation was performed using our existing implicit code \cite{kempf_2013}.

\begin{figure}
		\centering
		\includegraphics[width=\textwidth]{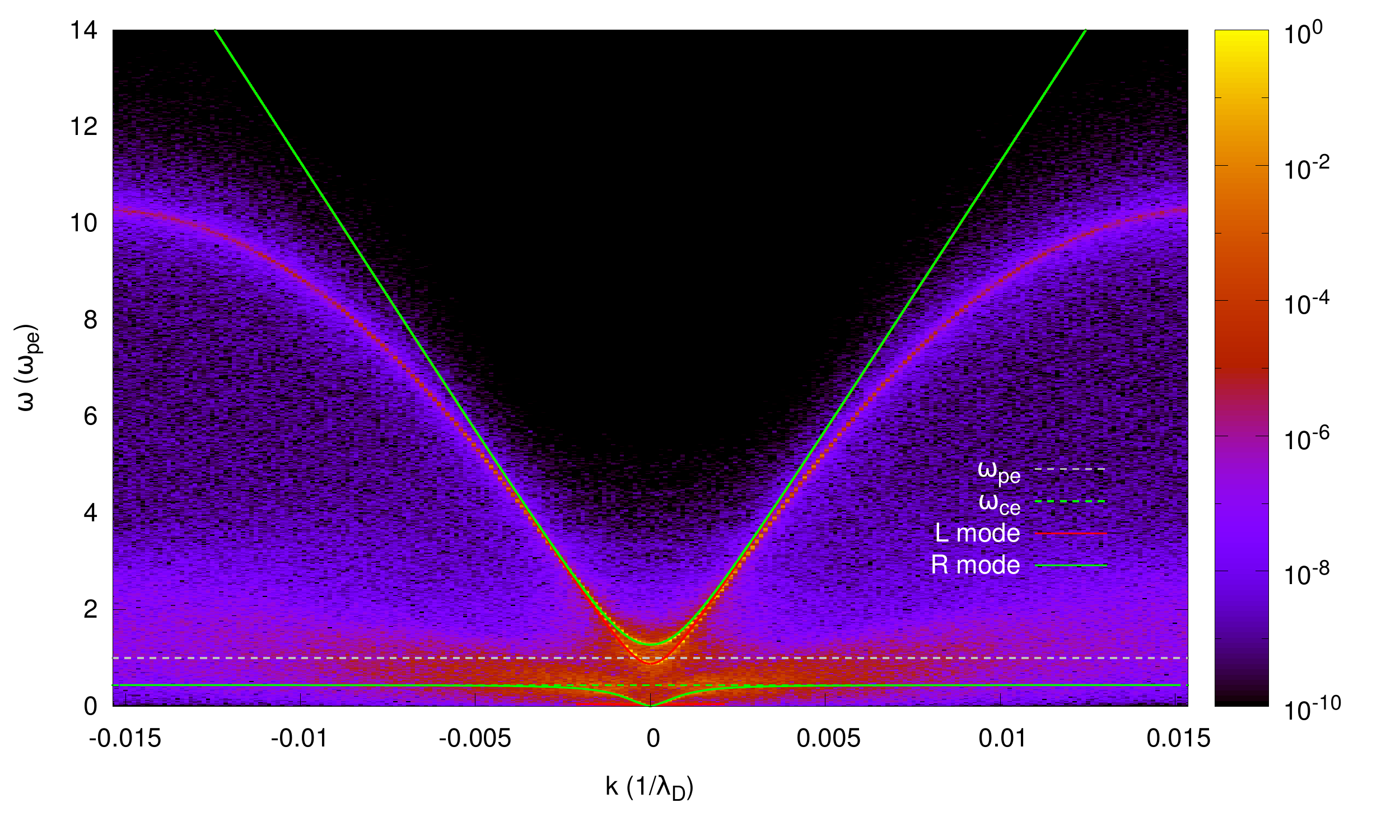}
		\caption{
			$E_\mathrm{y}(k_\mathrm{x}, \omega)$ dispersion plot for new implicit code.
			For this simulation, the charge density was smoothed spatially. $\Theta = 0.5$.
		}
		\label{fig:waves-norm}
	\end{figure}

\begin{figure}
		\centering
		\includegraphics[width=\textwidth]{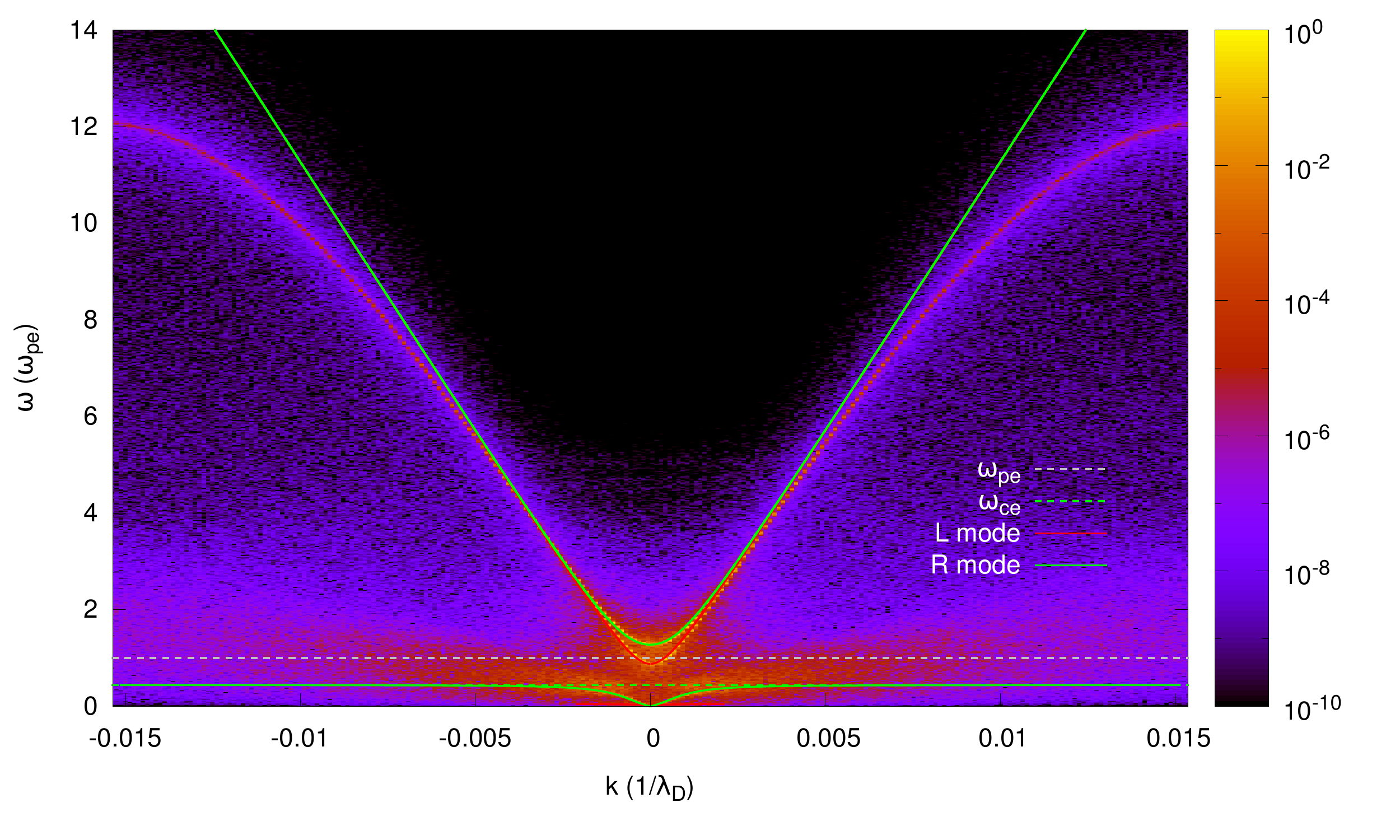}
		\caption{$E_\mathrm{y}(k_\mathrm{x}, \omega)$ dispersion plot for SOIDBERG.}
		\label{fig:waves-soid}
	\end{figure}

\begin{figure}
		\centering
		\includegraphics[width=\textwidth]{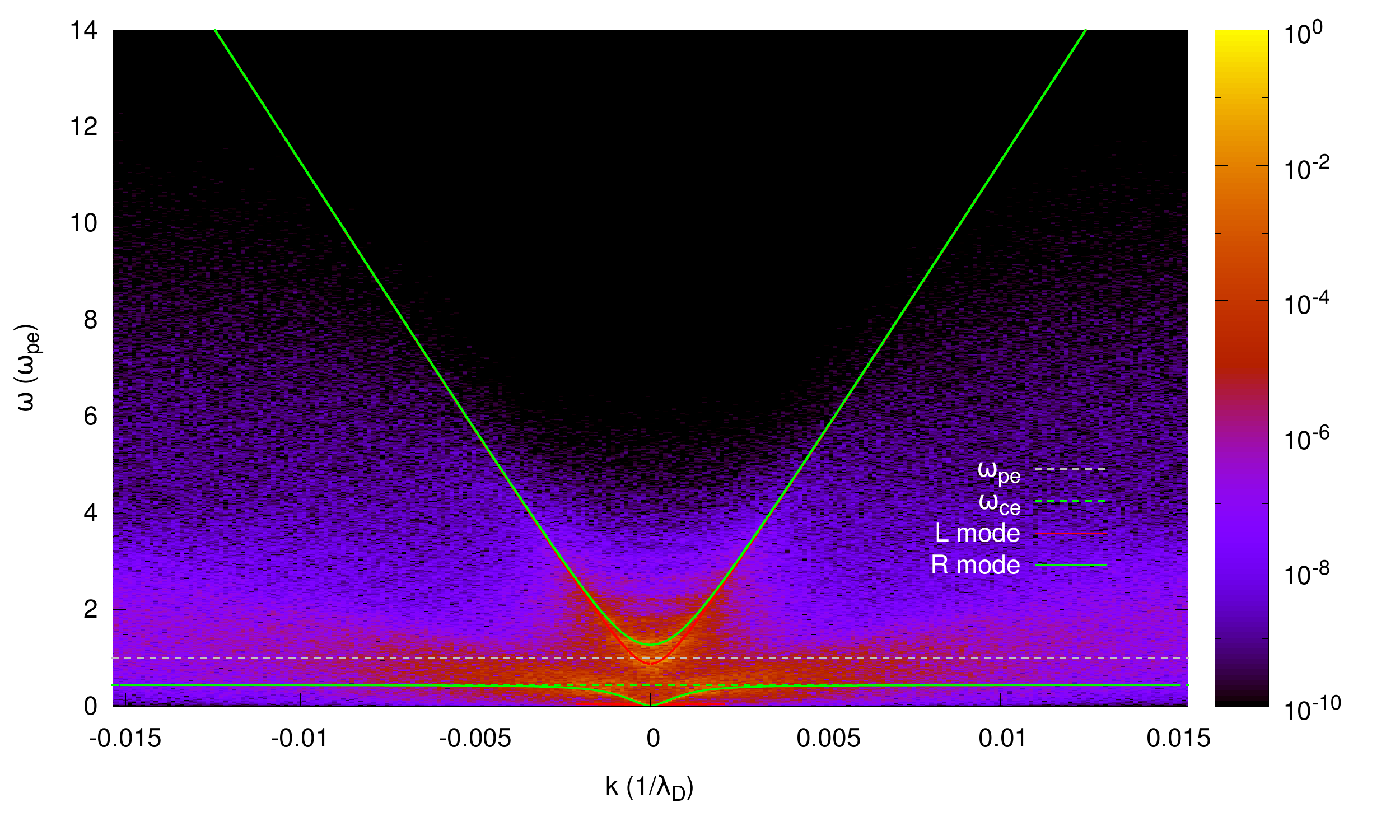}
		\caption{
			$E_\mathrm{y}(k_\mathrm{x}, \omega)$ dispersion plot for new implicit code.
			No spatial smoothing.
			$\Theta = 1.0$.
		}
		\label{fig:waves-time}
	\end{figure}

	In fig. \ref{fig:waves-norm}, a dispersion plot of the y-component of the electric field is shown with curves representing the theoretical dispersion relations overlaid.
	As can be seen, the wave modes are reproduced properly, with the electromagnetic mode showing a characteristic resonance at high $k$, owing to the finite grid used in PiC codes.
	A comparison to fig. \ref{fig:waves-soid} shows very similar behavior of the two codes.
	Notably, the new field solver leads to a EM-mode resonance at slightly lower $\omega$.
	Fig. \ref{fig:waves-time} demonstrates temporal filtering with $\Theta=1.0$.
	Clearly, the electromagnetic wave is almost completely removed from the simulation.
	Other wave modes are not affected, however, weak harmonics of the electron gyrofrequency are visible as noise.

	Energy conservation is not exact in this type of implicit PiC codes \cite{lapenta_kinetic} whereas in energy-conserving PiC codes, the momentum conservation is violated \cite{markidis_conserving}.
	Ideally, the total simulation energy decreases slowly since an increase might lead to instability.
	The $\Theta=0.5$ simulation and SOIDBERG both showed a decrease of about $1\%$.
	$\Theta=1.0$ led to an energy loss of about $0.4\%$.

	\begin{table}
		\centering
		\begin{tabular}{l|c|r}
			thermal velocity (electrons) & $v_\mathrm{th,e}$ & $0.05c$\\
			initial magnetic field & $\vec{B}_0$ & $(0.5, 0, 0)\,\mathrm{mT}$\\
			electron plasma frequency & $\omega_{\mathrm{pe}}$ & $2.0\cdot10^{8}\, \mathrm{rad/s}$\\
			Debye length & $\lambda_{\mathrm{D}}$ & $7.5\,\mathrm{cm}$ \\
			mass ratio& $m_{\mathrm{p}} / m_{\mathrm{e}}$ & 10\\
			timesteps & $N_\mathrm{t}$ & 4000\\
			length of timestep & $\Delta t$ & $4.1 \cdot 10^{-10}\,\mathrm{s}$ \\
			simulation volume & $N_\mathrm{x} \times N_\mathrm{y} \times N_\mathrm{z}$ & $256 \times 16 \times 16$\\
			cell edge length & $\Delta x$ & $26\,\mathrm{cm}$
		\end{tabular}
		\caption{Parameters of the wave dispersion simulations.}
		\label{tab:parameters-waves}
	\end{table}


\subsection{Filamentation instability}
	Instabilities are another common phenomenon sensitive to particle-wave interactions.
	The filamentation instability in particular is one of the earlier applications of PiC codes.
	Above all, the generation of strong magnetic fields is a good indicator of whether the code performs as expected \cite{Kilian_2011}.

	For demonstration purposes, we considered a fully three-dimensional and relativistic filamentation instability.
	The parameters for the two counter-streaming particle distributions are given in table \ref{tab:parameters-fila}.

\begin{figure}
		\centering
		\includegraphics[width=\textwidth]{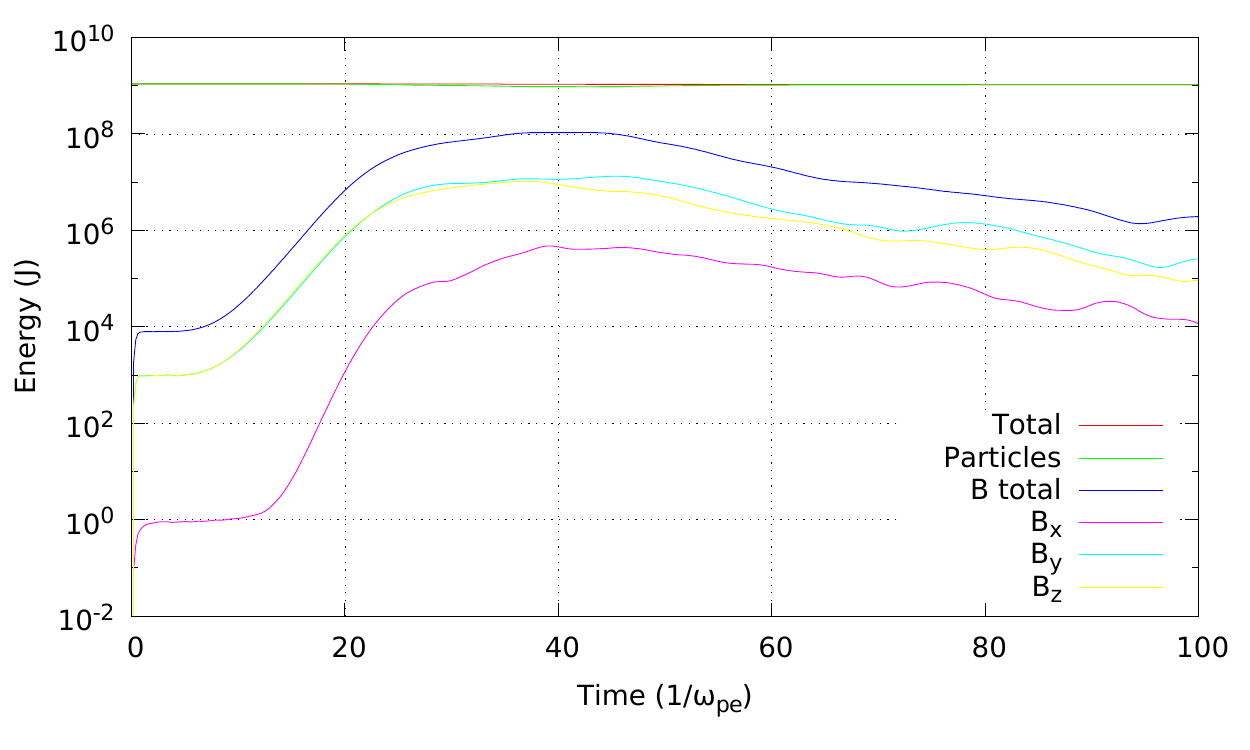}
		\caption{Energy output of the filamentation simulation.}
		\label{fig:fila-energy}
	\end{figure}

\begin{figure}
		\centering
		\includegraphics[width=.45\textwidth]{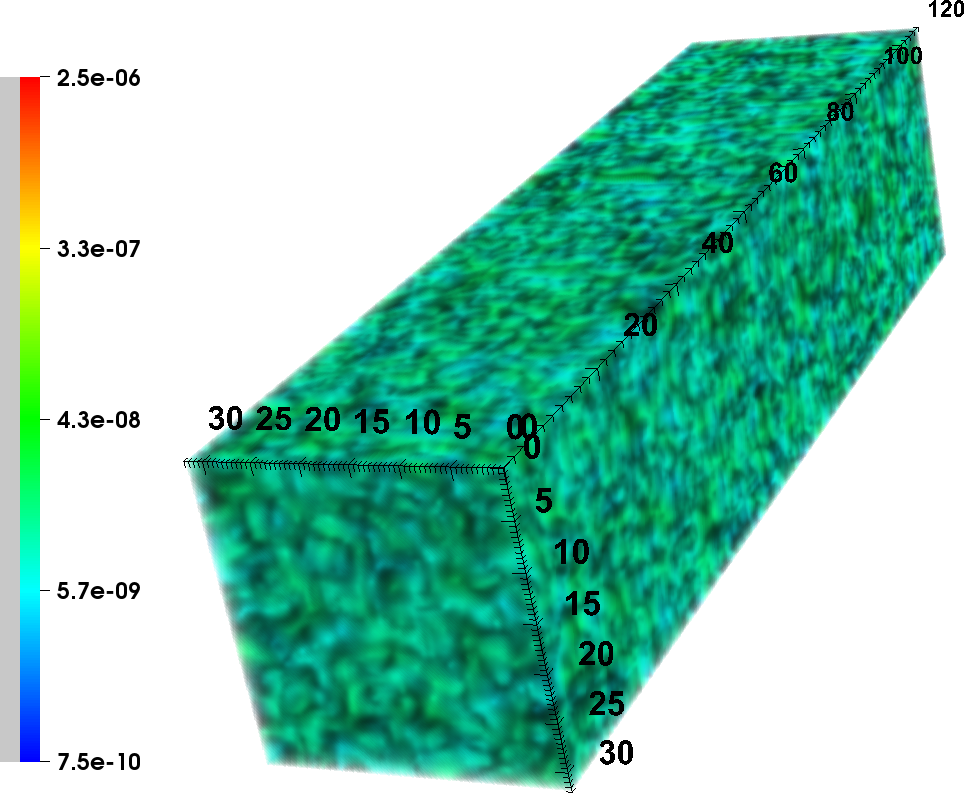}
		\includegraphics[width=.45\textwidth]{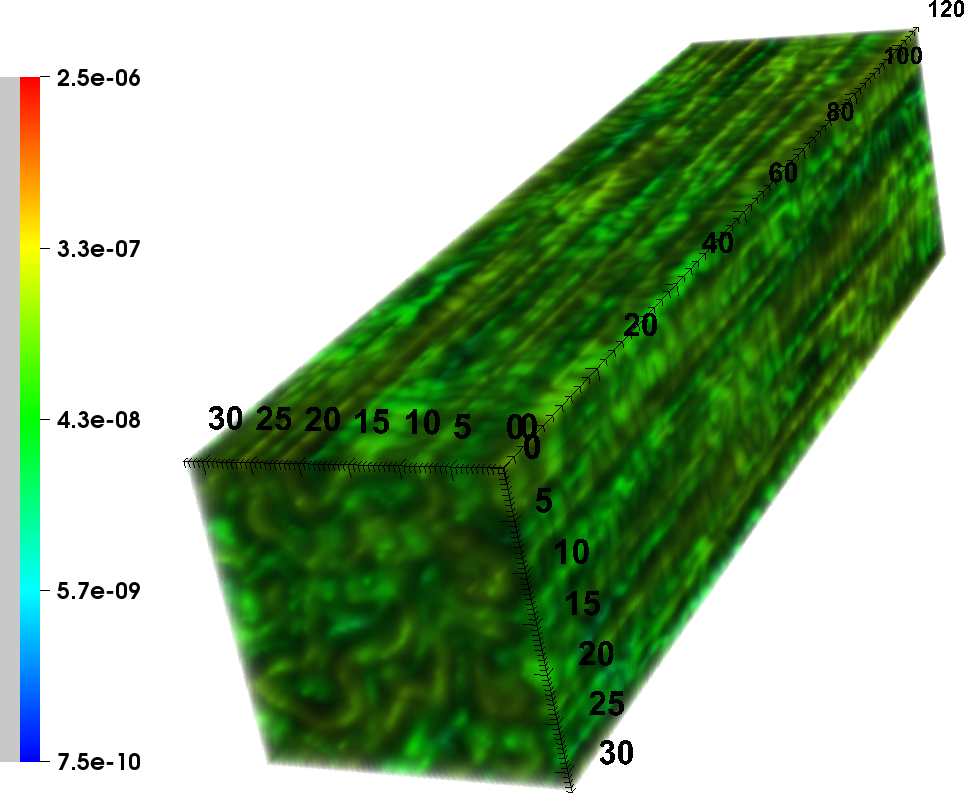}\\
		\includegraphics[width=.45\textwidth]{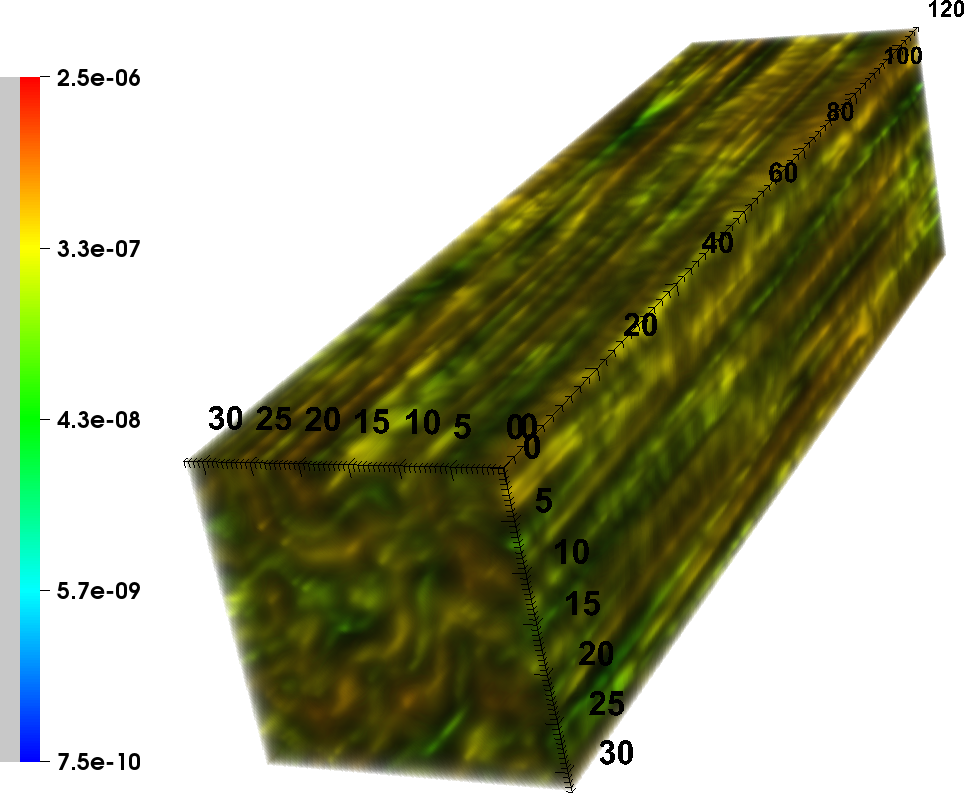}
		\includegraphics[width=.45\textwidth]{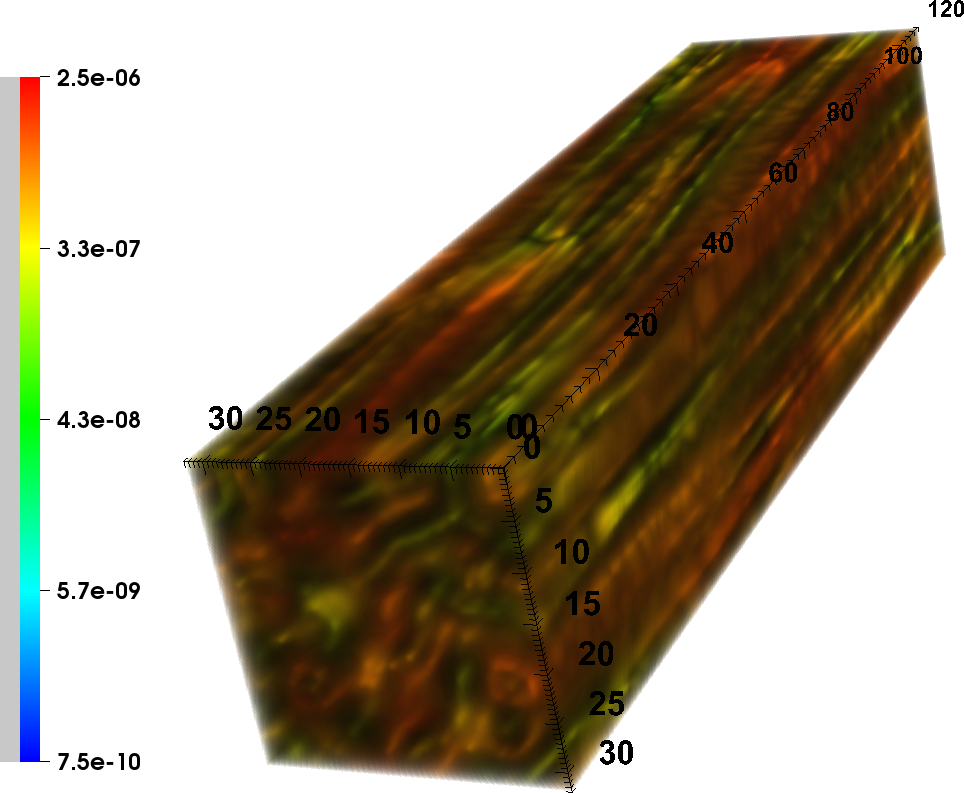}\\
		\includegraphics[width=.45\textwidth]{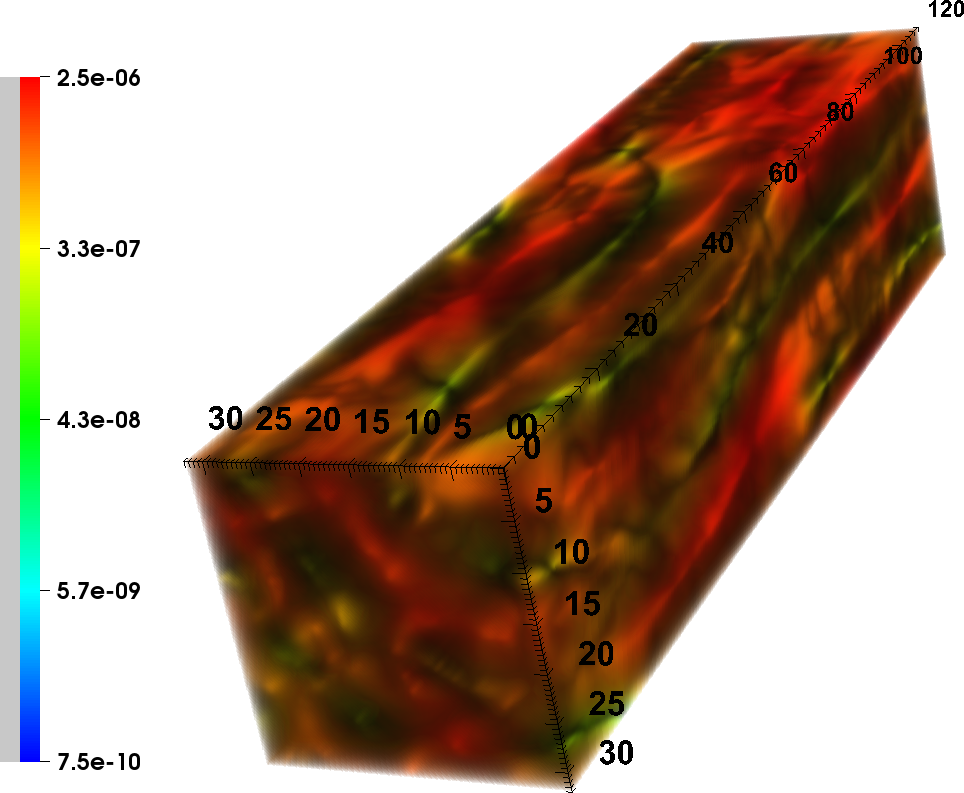}
		\includegraphics[width=.45\textwidth]{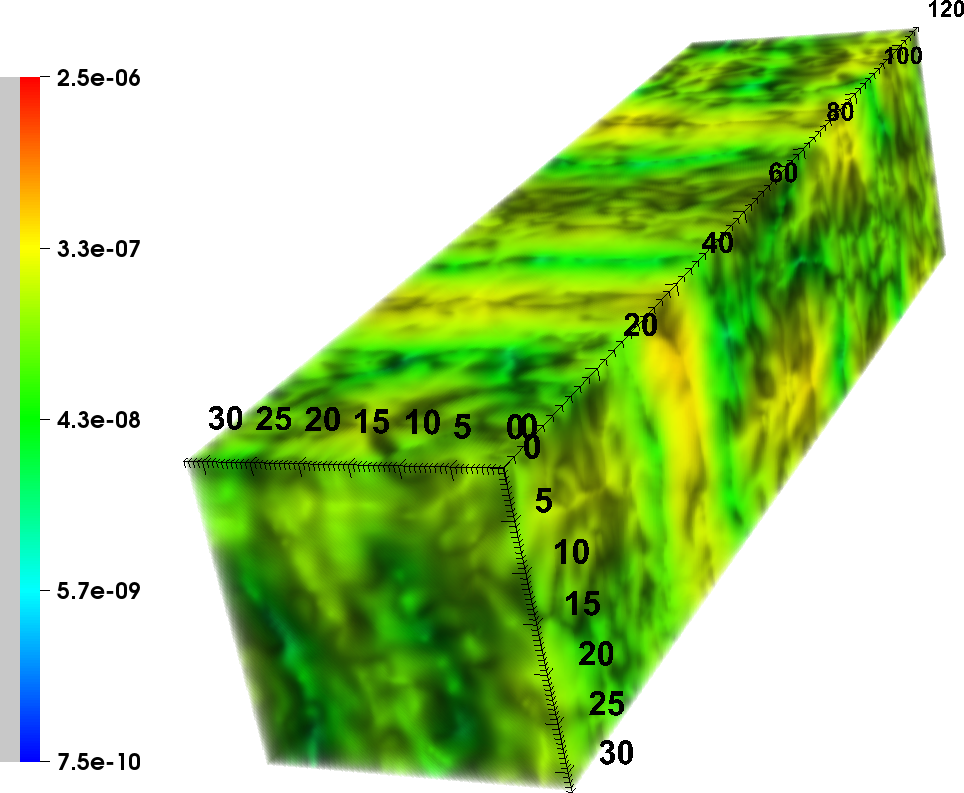}
		\caption{
			Time series of volumetric plots of the perpendicular magnetic field $\sqrt{B_\mathrm{y}^2+B_\mathrm{z}^2}$ for the filamentation simulation.
			$T = 5/\omega_\mathrm{pe}$, $T = 15/\omega_\mathrm{pe}$, $T = 20/\omega_\mathrm{pe}$, $T = 25/\omega_\mathrm{pe}$, $T = 35/\omega_\mathrm{pe}$, $T = 100/\omega_\mathrm{pe}$ (from left to right, top to bottom).
		}
		\label{fig:fila-plots}
	\end{figure}

	Instabilities generally proceed in an exponential fashion.
	In this case, the magnetic field perpendicular to the streaming direction should increase exponentially \cite{schlickeiser}.
	From the energy output of the simulation, fig. \ref{fig:fila-energy}, it is clear that an instability develops.
	Volumetric plots of the perpendicular magnetic field $\sqrt{B_\mathrm{y}^2+B_\mathrm{z}^2}$ (fig. \ref{fig:fila-plots}) show the rapid evolution of spatial structures.
	Out of a random magnetic field created by thermal fluctuations, filaments develop parallel to the streaming direction.
	Over the course of the simulation these filaments merge, forming larger structures and effecting significant perpendicular magnetic field amplitudes.
	After the instability reaches a saturation point, these strong fields decay slowly.

\begin{figure}
		\centering
		\includegraphics[width=\textwidth]{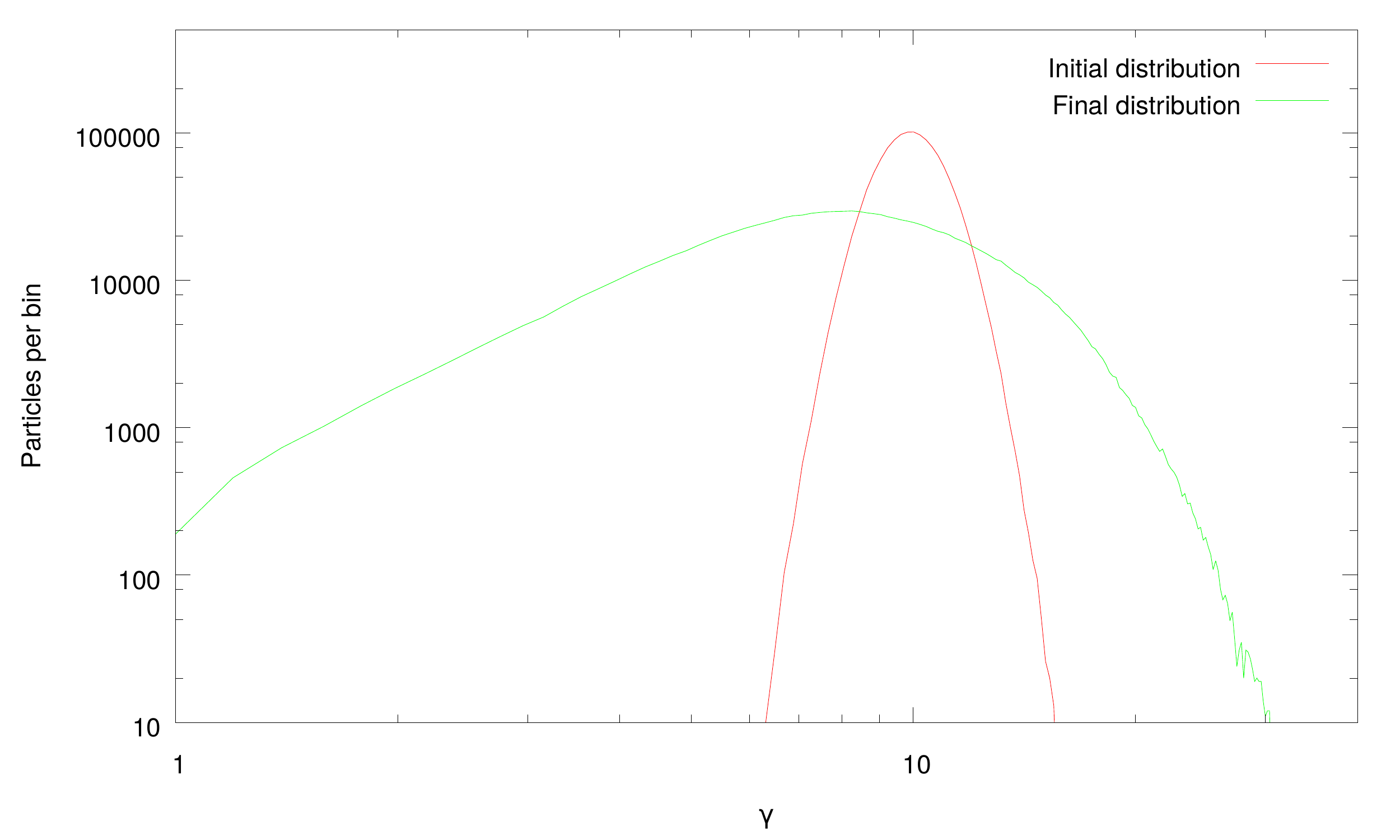}
		\caption{Initial and final electron energy distribution for the filamentation simulation.}
		\label{fig:fila-histogram}
	\end{figure}

	In addition to the the creation of strong fields, particle acceleration is expected to occur \cite{instability_accel}.
	For a histogram of the electron energy distribution before and after the simulation see fig. \ref{fig:fila-histogram}.
	Initially Maxwellian, the particle energy forms a high energy tail during filamentation.
	Most particles' kinetic energy, however, is converted to magnetic field energy.

	\begin{table}
		\centering
		\begin{tabular}{l|c|r}
			thermal velocity (electrons) & $v_\mathrm{th,e}$ & $0.1c$\\
			stream velocity & $v$ & $\pm 0.995c$\\
			electron plasma frequency & $\omega_{\mathrm{pe}}$ & $1.0\cdot10^{5}\, \mathrm{rad/s}$\\
			Debye length & $\lambda_{\mathrm{D}}$ & $3.0 \cdot 10^4 \,\mathrm{cm}$ \\
			mass ratio& $m_{\mathrm{p}} / m_{\mathrm{e}}$ & 2\\
			timesteps & $N_\mathrm{t}$ & 500\\
			length of timestep & $\Delta t$ & $2.0 \cdot 10^{-6}\,\mathrm{s}$ \\
			simulation volume & $N_\mathrm{x} \times N_\mathrm{y} \times N_\mathrm{z}$ & $128 \times 32 \times 32$\\
			cell edge length & $\Delta x$ & $1.1 \cdot 10^5\,\mathrm{cm}$
		\end{tabular}
		\caption{Parameters of filamentation simulation.}
		\label{tab:parameters-fila}
	\end{table}


\section{Concluding remarks}
	A production-ready and, above all, simple implementation of the relativistic moment implicit particle-in-cell algorithm was presented.
	Utilizing advanced techniques like GMRES and Newton-Krylov, PICPANTHER is well-equipped for the simulation of difficult plasma physics problems.
	Moreover, the concise and straightforward style allows for easy understanding and expansion of the code.

	Several test cases were examined and the code's properties determined.
	The simulations accurately reflect the physical processes, even in extreme cases like the filamentation instability.
	Performance was found to be decent and scalability tests on SuperMUC suggest good weak scaling behavior.

\section{Acknowledgments}
	The authors gratefully acknowledge the Gauss Centre for Supercomputing e.V. (www.gauss-centre.eu) for funding project \textit{pr45ye} by providing computing time on the GCS Supercomputer SuperMUC at Leibniz Supercomputing Centre (LRZ, www.lrz.de).
	AK acknowledges support by grant Schl 201/23-1 within the priority programme 1573: Physics of the Interstellar Medium of the Deutsche Forschungsgemeinschaft.
	PK acknowledges support by the Max-Planck Princeton Center for Plasma Physics (MPPC) and CRC 963 Astrophysical Flow Instabilities and Turbulence of the Deutsche Forschungsgemeinschaft.
	UG acknowledges support by the Deutsche Forschungsgemeinschaft through grant GA1968/1-1.

\clearpage





\bibliographystyle{model1-num-names}
\bibliography{paper}







\end{document}